\documentclass[aps,twocolumn]{revtex4}

\usepackage{dcolumn}
\usepackage{bm}
\usepackage{latexsym}
\usepackage{graphics}

\begin{document}
\title{Slow beams of massive molecules}
\author{Sarayut Deachapunya$^{1,2}$, Paul J. Fagan$^3$,  Andr\'{a}s G. Major$^1$,
Elisabeth Reiger$^4$, Helmut Ritsch$^5$, Andr\'{e} Stefanov$^1$,
Hendrik Ulbricht$^1$, and Markus Arndt$^1$} \affiliation{$^1$
Faculty of Physics, University of Vienna,
Boltzmanngasse 5, A - 1090 Vienna, Austria\\
$^2$ Department of Physics, Faculty of Science, Burapha
University, Chonburi 20131, Thailand\\
$^3$ R\&D The DuPont Company, PO Box 80328, Experimental Station, Wilminton, DE 19880-0328, USA\\
$^4$ Kavli Institute of Nanoscience, TU Delft , Lorentzweg 1, NL-2628 CJ Delft, The Netherlands\\
$^5$ Institute of Theoretical Physics, University of Innsbruck,
 Technikerstra{\ss}e 20, A-6020 Innsbruck, Austria.}

\date{\today}
\begin{abstract}
Slow beams of neutral molecules are of great interest for a wide
range of applications, from cold chemistry through precision
measurements to tests of the foundations of quantum mechanics.  We
report on the quantitative observation of thermal beams of
perfluorinated macromolecules with masses up to 6000\,amu, reaching
velocities down to 11\,m/s. Such slow, heavy and neutral molecular
beams are of importance for a new class of experiments in
matter-wave interferometry and we also discuss the requirements for
further manipulation and cooling schemes with molecules in this
unprecedented mass range.
\end{abstract}
%
%

\maketitle
\section{Introduction}
\label{intro} Stimulated by the great success of atom cooling and
trapping experiments~\cite{Varenna1993a,Weidemuller2003a}, much
effort has also been directed at demonstrating slow or cold
molecular beams.

This includes the pioneering work on the
deceleration~\cite{Bethlem2000b} and trapping in electric
~\cite{Crompvoets2001a,Junglen2004b} or magnetic
fields~\cite{Doyle1995a,Sawyer2007a}. Similarly the deceleration of
the heavier YbF to 287\,m/s was realized for new precision
experiments~\cite{Tarbutt2004a}. Optical dipole forces were employed
to slow C$_{6}$H$_{6}$ to 300\,m/s~\cite{Fulton2004a}. A
back-rotating nozzle was used for reducing the speed of SF$_{6}$ to
55 m/s ~\cite{Gupta2001a} and back-rotating silicon paddles
significantly decelerated fast Helium atoms ~\cite{Narevicius2007a}.
Laser implantation into cryogenically cooled Helium allowed the
formation of PbO beams at 40\,m/s ~\cite{Maxwell2005a}. The slowing
of NO molecules in billiard-like collisions with Ar reduced their
velocity to 15\,m/s~\cite{Elioff2003a}. Slow and cold dimers were
also formed in a reaction between counter-propagating H and halogen
atoms~\cite{Liu2007a}. Finally, cavity assisted optical manipulation
methods~\cite{Horak1997a} have recently  been proposed for cooling
external and internal~\cite{Morigi2007a} degrees of freedom in small
molecules.
\begin{figure}
\resizebox{1\columnwidth}{!}{
   \includegraphics{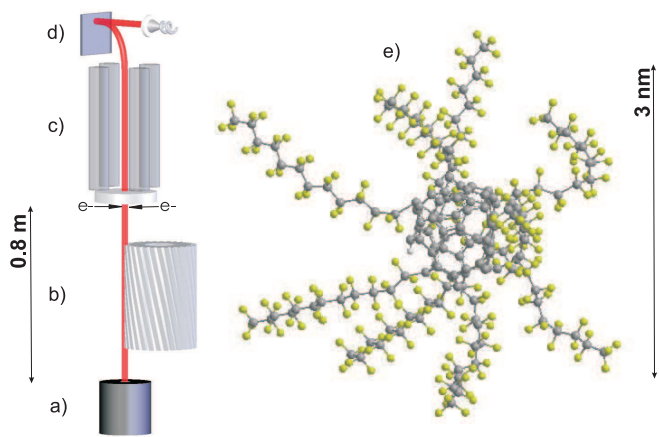}}
   \caption{Setup for sublimation and velocity measurements:  a) thermal source (not to scale), b) velocity selector, c) electron impact ionization quadrupole mass
   filter and d) detection unit. The structure of the perfluoroalkylated carbon sphere e), also designated as
PerfluoroC$_{60}$.} \label{Fig1:MoleculeShapes}
\end{figure}
The rapid evolution of molecule experiments opens the question if
some of the new methods could also be applied to distinctively more
massive systems. This is particularly interesting with regard to the
stringent requirements of quantum interferometry with massive
compounds~\cite{Arndt1999a}.

Common to all such experiments is the need to volatilize complex
materials at sufficiently low kinetic energy, which is usually a
great challenge. For organic molecules one often observes an
increase in the particle's electric polarizability, dipole moment
and number of weak bonds when the number of atoms per molecule is
augmented. Correspondingly, there is an overall trend for large
molecules to have a low vapor pressure and a high fragmentation
probability at elevated temperatures.

Some of these problems can be circumvented by recurring to matrix
assisted laser desorption (MALD)~\cite{Tanaka1988a}, jet expanded
laser desorption (JETLD)~\cite{Grotemeyer1986a} or electro-spray
ionization (ESI)~\cite{Fenn1989a}, but only at the expense of
producing either fast (MALD,JETLD) or highly charged (ESI) molecular
beams.
\begin{figure*}
\resizebox{1.8\columnwidth}{!}{
\includegraphics{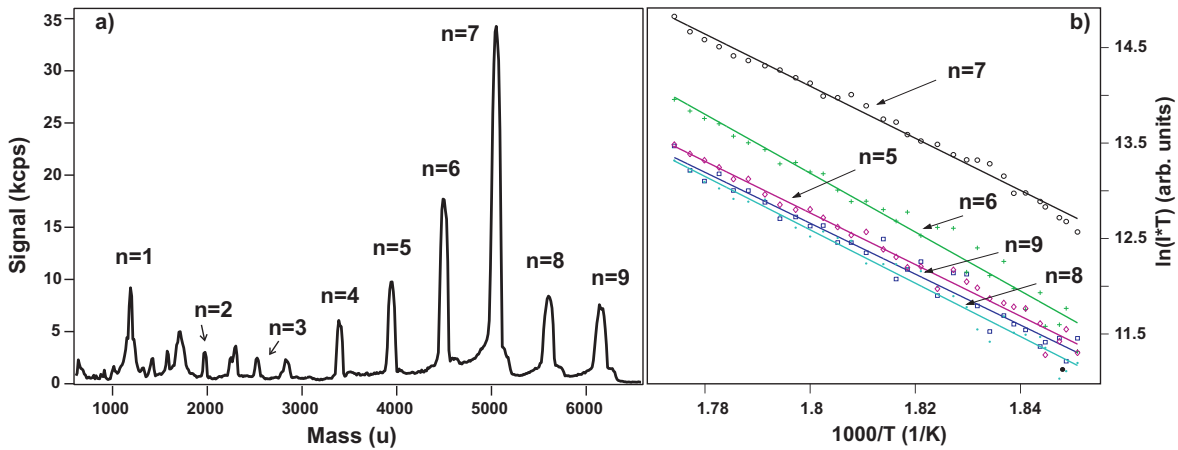}}
\caption{a) shows a typical mass spectrum of PerfluoroC$_{60}$ at
545\,K. The integer $n$ counts the number of perfluoroalkylated
side-chains attached to the central buckyball. The mass spectrometer
parameter are optimized for most efficient detection of the $n=7$
compound. b) is the Arrhenius plot to evaluate the sublimation
enthalpy for n=5...9 compounds.
 \label{Fig2:perfluoroC60Sublimation}}
\end{figure*}

\section{Experimental characterization of large molecules in the gas phase}
\label{sec:2} In marked contrast to these observations, we here
report on new perfluoroalkylated neutral particles in the mass range
up to more than 6000\,amu, whose vapor pressures are sufficiently
high and whose velocities are sufficiently low to open a new
experimental window for new coherent (interferometry, molecular
lenses) or incoherent (cooling) molecular manipulation schemes.

\subsection{Molecule}
In particular we study the perfluoroalkylated carbon-sphere
C$_{60}$[(CF$_{2}$)$_{11}$CF$_{3}$]$_{n}$H$_{m}$, where $m \in
\{0...2 \}$ is the number of attached H atoms and $n\in\{0...9 \}$
is the number of fluorinated side-chains attached to the $C_{60}$
core. Figure~\ref{Fig1:MoleculeShapes} shows an energetically
non-relaxed view of the molecule, to illustrate its overall
structure and complexity. The molecules~\cite{Fagan1993a} were
synthesized by one of the authors (P F). The mass of a nanosphere
with n=9 side-chains exceeds that of any molecule in all previous
slow-beam studies by more than an order of magnitude, and even that
of a small protein, such as insulin (m$\sim$5700\,amu).

\subsection{Beam machine}
Our experiments were performed in a vertical fountain configuration,
which is crucial for measuring high intensities in particular for
slow molecules, see Fig.~\ref{Fig1:MoleculeShapes}. The material was
evaporated in a furnace with a circular aperture of 500 $\mu$m
diameter. The home-made helical velocity selector, shown in
Fig.~\ref{Fig1:MoleculeShapes} is located 5 mm above the furnace
aperture and has a length of 140 mm and a radius of 48 mm. The
velocity selection is done by channels (grooves) in the rim of the
selector which have a slope with regard to the selector axis. The
angle of between the grooves and the rotor axis is 13 rad. The
angular velocity of the rotor determines the mean velocity of the
transmitted molecules. The bandwidth of $\Delta v/v=5\%$ (FWHM) is
determined by the aspect ratio (width/length) of the milled grooves
in the selector. For our model, a rotation frequency of 1~Hz
corresponds to a molecule velocity of 1.08~m/s and the mean
transmitted velocity scales linearly with the rotor frequency. The
phase stability of the rotation of the selector was carefully
checked and maintained with a stroboscopic flash lamp operating at
the rotation frequency with $\Delta f = 0.1~\%$. Having passed the
selector, and after a drift region of 0.5 m above the furnace the
perfluorinated compounds were finally detected using electron impact
ionization quadrupole mass spectroscopy in a differentially pumped
second vacuum chamber. We used an Extrel quadrupole mass
spectrometer with electron energy of E$_{\rm kin}$=40..70\,eV which
allows to detect molecules with masses up to 9000\,amu. We here
report on the first on electron impact ionization for detecting
those large perflouroalkylated molecules~\cite{Fagan1993a}. For the
mass spectrum shown in Figure\,\ref{Fig2:perfluoroC60Sublimation}~a)
the spectrometer was optimzed to the signal of nanospheres with n=7
side-chains (see discussion below).

\subsection{Characterization of the carbon nanospheres}\label{sec:5}
Figure\,\ref{Fig2:perfluoroC60Sublimation}~a) shows a mass spectrum
of the post-ionized perfluoroalkylated carbon spheres. We identify
molecules with between one and nine intact side chains. The relative
intensity of these peaks is determined by the chemical synthesis and
might be influenced by the chemical rearrangements during long-time
storage, spectrometer settings as well as fragmentation processes
during both the evaporation and the ionization process. But the
variation of their peak height with the furnace temperature is an
absolute measure for their sublimation enthalpy. In order to
quantify this, we linearly ramp the temperature with a heating rate
of 0.7~K/min and record the exponential increase of the count rate.
Using the Clausius-Clapeyron equation we evaluate the sublimation
enthalpy $\Delta H_{sub}$ from an Arrhenius fit to the data (see
Figure\ref{Fig2:perfluoroC60Sublimation}~b). The results are
summarized in Table~\ref{tableI}.

We have also estimated the scalar polarizability $\alpha$ using the
software Gaussian03~\cite{Gaussian03a} with a reduced Hartree-Fock
method and the 6-31~G polar basis. We find a scalar static
polarizability of $\alpha=194~\AA^3$ and a permanent dipole moment
of about 6 Debye for the carbon nanosphere with n=7. The compound
with n=1 was calculated to have a static polarizability of
$\alpha=84~\AA^3$ which is close to the measured value for an
individual $C_{60}$ molecule~\cite{Berninger2007a}. Each additional
side-chain adds to the total polarizability with about $18~\AA^3$.
The low $\alpha/m$-value is consistent with the unusually high vapor
pressure of these perfluorinated compounds, making them particularly
useful for generating slow thermal beams.

All sublimation enthalpies are equal within their error bars. This
is both compatible with an possible initial mixture of different
perfluoroalklated carbonspheres with very similar $\alpha$/m ratios,
as well as with a monodisperse distribution of large molecules,
undergoing fragmentation in the ionization process.
\begin{table}
\caption{Sublimation enthalpies for large
perfluoralky-functionalized molecules: The temperature interval for
the sublimation studies (T$_m$) and the experimental decomposition
temperature (T$_{d}$) were T$_m$=540-563\,K and T$_d$=650\,K. Error
bars are evaluated from curve fitting deviations.} \label{tableI}
\begin{tabular}{lllll}
\hline\noalign{\smallskip}
Molecule &  Mass (u) & $\Delta H_{sub}$[kJ/mol] \\
\noalign{\smallskip}\hline\noalign{\smallskip}
PerfluoroC$_{60}$, n=9 & 6291 &  217 $\pm $ 15 \\
PerfluoroC$_{60}$, n=8 & 5672 &  227 $\pm $ 13 \\
PerfluoroC$_{60}$, n=7 & 5053 &  222 $\pm $ 8 \\
PerfluoroC$_{60}$, n=6 & 4434 &  251 $\pm $ 16 \\
PerfluoroC$_{60}$, n=5 & 3815 &  220 $\pm $ 11 \\
\noalign{\smallskip}\hline
\end{tabular}
\end{table}
The compound of the PerfluoroC$_{60}$ with seven side chains had the
highest absolute signal in this series of experiments. It reached up
to 750000\,cps at 635\,K, corresponding to a molecular flux of
$10^{11}$\,s$^{-1}$cm$^{-2}$ in the detection region - that is 800
mm above the furnace. From the measured flux and the velocity we can
calculate the number density to be $10^{13}$\,cm$^{-3}$ for a
distance 3~mm above the source exit (see cavity focusing below).
Here we assume a rather conservative total detection efficiency for
the neutral particles of about $\eta=10^{-4}$. This estimate is
based on the observation by Bart et al. \cite{Bart2002a} that the
electron impact ionization cross section per bond can be
extrapolated to larger perfluoinated hydrocarbons. The electron
impact ionization cross section here was thus estimated to be
$\sigma_{PerfluoroC_{60}} = 2.7 \times 10^{-18}$\,m$^2$.

The observation of a rather high flux of intact neutral and very
massive molecules is a key result of great importance for the
proposed interferometry and optical manipulation applications.

\subsection{Velocity measurments}
\begin{figure}
\resizebox{1\columnwidth}{!}{
    \includegraphics{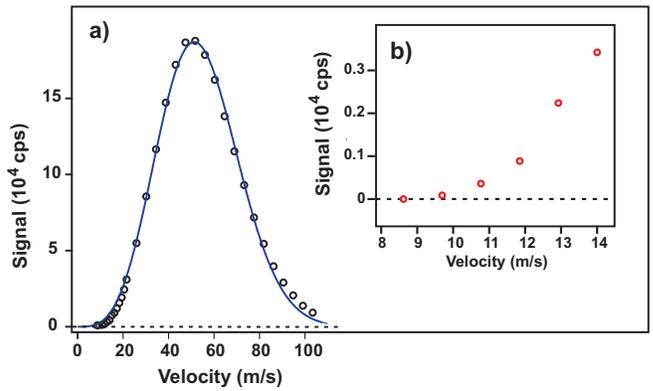}}
    \caption{a) Velocity distribution of the isomer n=7 (m = 5053\,amu)
    of the perfluoroalkylated carbon nanosphere at 585\,K. The solid
    line represents a fit with a floating Maxwell-Boltzmann
    distribution. b) The inset shows molecules even at velocities down
to 11\,m/s. \label{VelocityPerfluroC60}}
\end{figure}

Earlier experiments already showed that matter wave interferometry
is well feasible for de Broglie wavelengths larger than about one
picometer~\cite{Arndt1999a}. For molecules with a mass of 5000\,amu
this limit is touched at a velocity of 55\,m/s. Help in this
situation may come from the expectation that massive molecules from
thermal sources propagate at low mean velocities. We therefore
present in Fig.~\ref{VelocityPerfluroC60} the velocity spectrum of
the carbon nanospheres with  $n=7$ side chains. We observed a
floating Maxwell-Boltzmann distribution
\begin{equation}\label{maxwell}
    f(v) =  v^2 \exp(-\frac{m(v-v_m)^2}{2 k_B T}) ,
\end{equation}
with a most probable velocity of v$_m$=51\,m/s, which is about 15\%
faster than $v_{m}=\sqrt{2k_BT/m}=44$\,m/s, which would be expected
for a fully effusive beam. The width of the curve fits to an
ensemble temperature of 302\,K, compared to the source temperature
of 585\,K. This indicates that the source operates at the transition
between an effusive and a weakly supersonic beam~\cite{Scoles1988a}
which provides already translational cooling by a factor of two. But
even more importantly, we still find a detectable fraction of
particles at velocities down to 11\,m/s. Such low velocities may
come as a surprise, given the well-established fact that in atomic
fountains the slow fraction of the atomic ensemble is usually
suppressed by collisions with faster particles~\cite{Ramsey1990a}.

Although while we also see a shift to higher velocities, we still
maintain an overall slow envelope, since our seed gas is the
molecule itself with a mass of about 5000 amu and the experimental
observation is in good agreement with elementary theoretical
expectations.

The rather significant signal at low velocities is very promising
for testing matter-wave physics in an unprecedented mass range using
a new interferometer concept that has recently been developed in our
group~\cite{Gerlich2007a}. With regard to future matter wave
experiments it is important to see, that the geometrical cross
section of perfluoralkylated particles exceeds that of C$_{60}$
already by a factor of about ten, while their scattering cross
section, determined by the van der Waals interaction, remains still
comparable. This is particularly relevant with respect to the
suppression of collisional decoherence~\cite{Hornberger2003a}.
\begin{figure*}
\resizebox{2\columnwidth}{!}{
\includegraphics{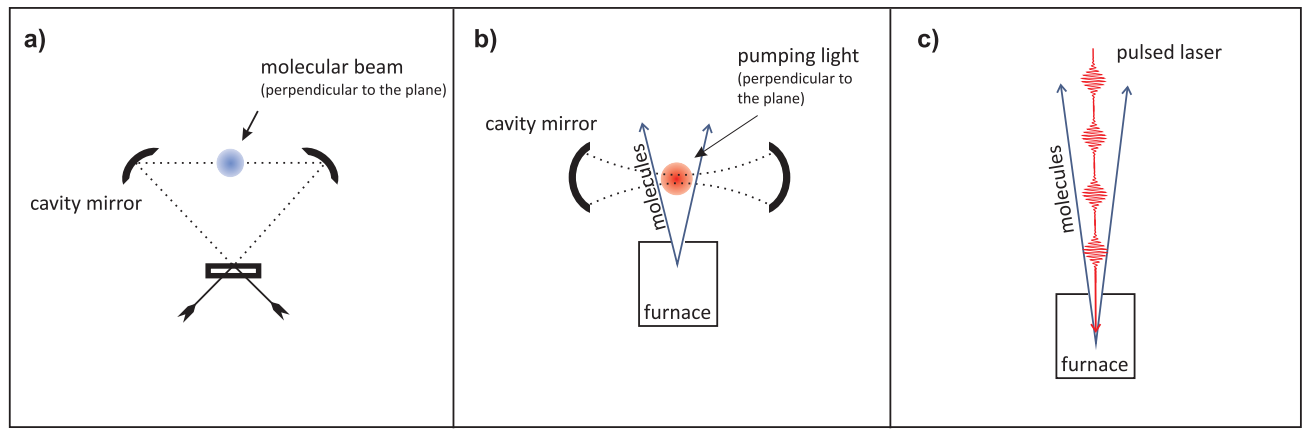}}
\caption{Discussed setups: A ring-cavity for optical
focusing/collimation (a) and a linear cavity for optical cooling (b)
of molecules, generated form a thermal source. The pump laser beam
is perpendicular to both the axis of the cavity and the direction of
the molecular beam for the linear cavity. Only the light scattered
by the molecules couples into the cavity. (c) is a setup for
molecular beam slowing with counter-propagating pulsed laser light.}
\label{setup_cavity}
\end{figure*}

\section{Optical manipulation of massive molecular beams}
Given the low velocity of these heavy molecules it is intriguing to
also explore the possibilities for new post-processing schemes to
reshape and possibly further increase the phase-space density using
off-resonant light fields. We start by first estimating the laser
power which is required to manipulate molecules with a thermal
kinetic energy of $\sim$50\,meV. This corresponds to a velocity of
44 m/s for the n=7 perfluoroalkalyted nanosphere. A far-detuned
Gaussian laser beam of power $P$, focused to a waist of
$w_{0}$=100$\,\mu$m, creates a dipole potential of well-depth
\begin{equation}\label{potential}
     U = \frac{2\alpha P}{(\varepsilon_0 c \pi w_0^2)},
\end{equation}
i.e. of 3.3\,neV per Watt for a molecular polarizability of 200$\AA
^3$. A power of $P$=15\,MW is therefore needed, if we wish to fully
compensate the kinetic energy of our supermassive molecules. This
power can be provided by a common Q-switched laser with a pulse
energy of 75\,mJ delivered in a pulse duration of 5\,ns. However, we
also have to consider that about N$_{abs}$ photons are absorbed by
each molecule during the interaction time $\tau$ with the laser,
where
\begin{equation}\label{absorption}
    N_{abs} = \frac{I_0 \sigma \tau}{(h\nu)} = \frac{2P \sigma \tau}{(\pi
    w_0^2h\nu)} .
\end{equation}
\begin{figure*}
\resizebox{1.5\columnwidth}{!}{ \centering
\includegraphics{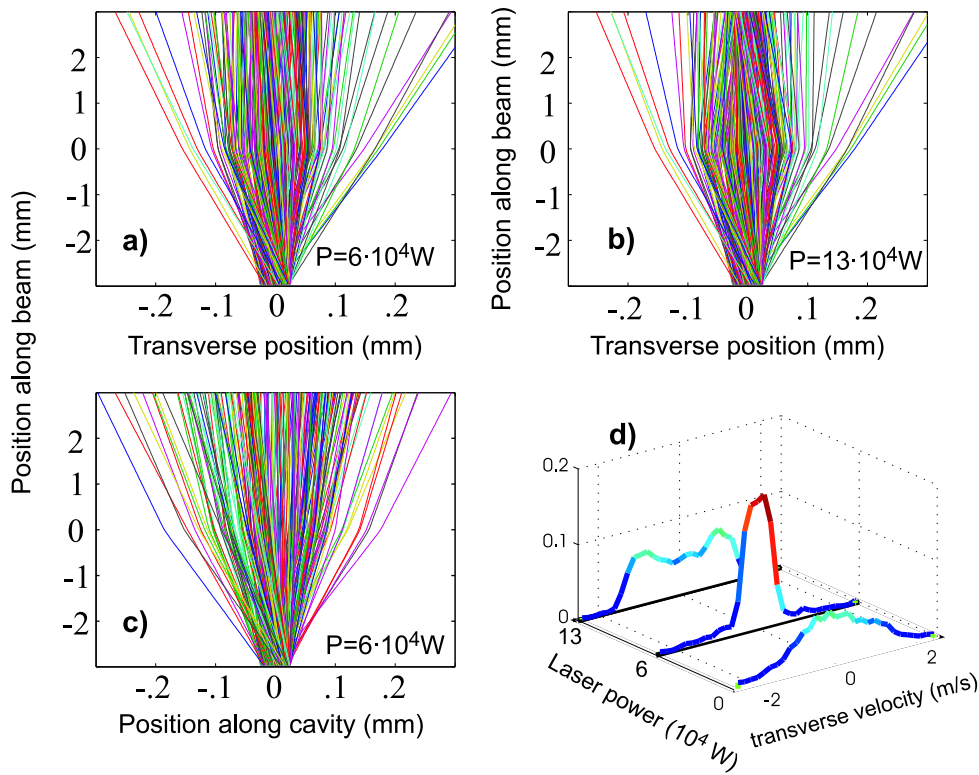}}
\caption{Numerical simulation of transverse optical focusing of a
molecular beam in a build-up ring cavity focused at the origin of
our coordinate system, 3\,mm above the source exit. Panels a) and b)
show focusing transverse to the laser and molecular beam for two
different intra-cavity laser powers, while panel c) shows the motion
along the laser beam. In panel d) we show the final molecular
velocity distributions after transit of the field for three
different intra-cavity powers.} \label{Transversefocussing}
\end{figure*}

Since our perfluorinated compounds have an NIR (1064\,nm) absorption
cross section of $\sigma _{1064}=3\times 10^{-23}$\,m$^2$, a 75\,mJ
stopping laser would deposit about 770 photons in each molecule and
would thus be likely to destroy the particle. The $\sigma$-value was
measured for solvated molecules and it might be smaller by an order
of magnitude for molecules in the gas-phase~\cite{Gotsche2007a}.

Higher intensities can be achieved with shorter laser pulses: Let us
consider a picosecond laser with a pulse duration of 7.5\,ps, a
pulse energy of 3\,mJ, a wavelength of 1064\,nm, and a waist of
1\,mm, which is counter-propagating with respect to the molecular
beam (see Figure~\ref{setup_cavity}~c). It would generate an
off-resonant slowing/focusing potential of 13\,meV in both the
transverse and the longitudinal molecular beam direction. A fraction
of the molecules will be decelerated by about 6\,m/s. Only 0.3 IR
photons would be absorbed by each molecule per light pulse. Slowing
of our molecules by pulsed laser light will therefore work as good
as reported before for benzene molecules~\cite{Fulton2004a}.
\begin{figure}
\resizebox{0.95\columnwidth}{!}{
\includegraphics{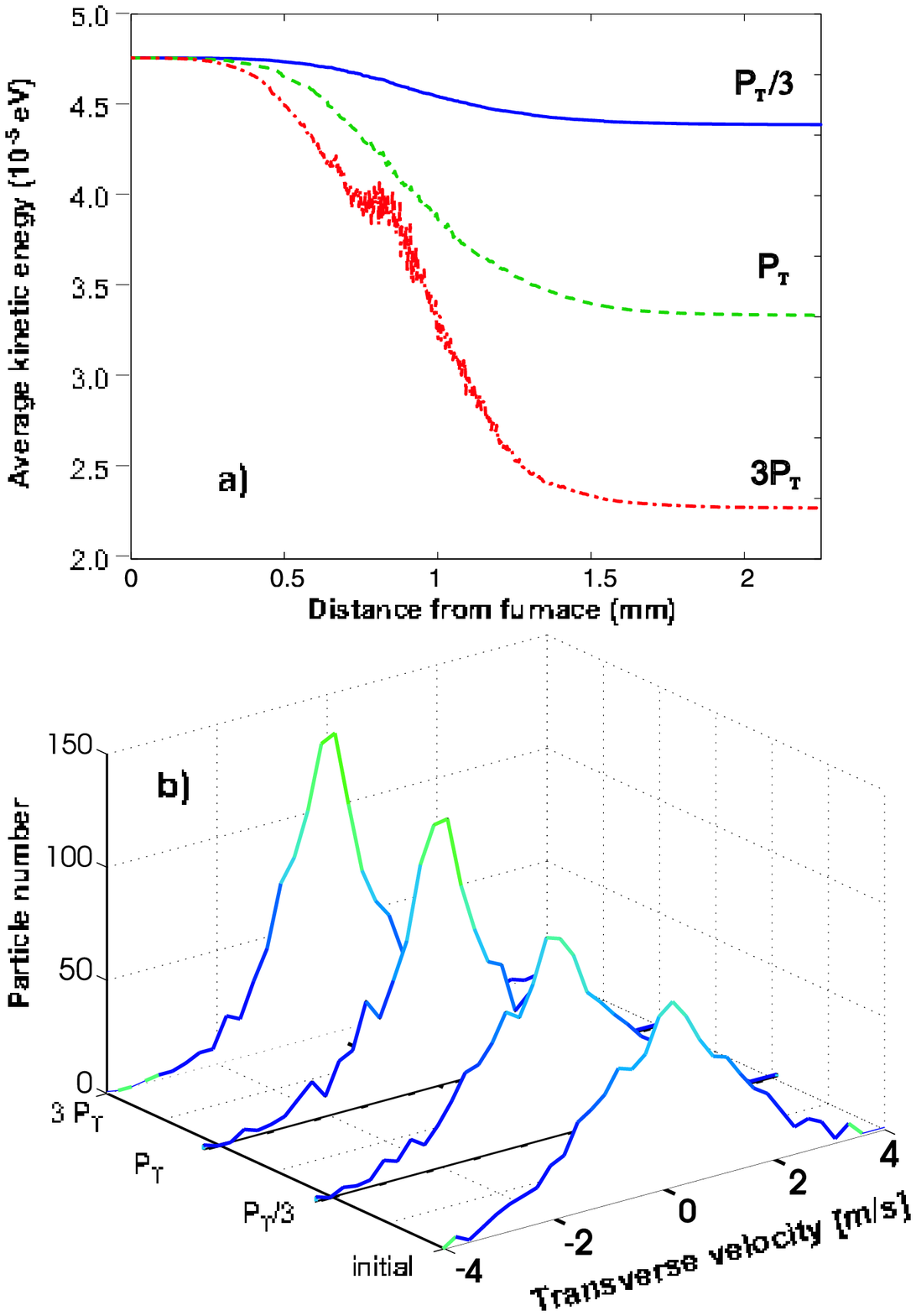}}
\caption{Cavity assisted optical cooling in one transverse molecular
beam direction: (a) Time evolution of kinetic energy along cavity
axis for $N=1000$ particles of mass $5000$\,amu, passing through a
cavity with waist $w_0=400\mu$m with average velocity of $10$\,m/s
($\Delta v = 1.5$\,m/s) for different laser powers $P=P_T/3, P_T,
3P_T$, where $P_T = 1$\,kW is the effective self-organization
threshold, (b) initial and final transverse velocity distributions
for these parameters. Note that we rescaled the interaction strength
to mimic the actually higher molecule density from the source.}
\label{Transversecooling}
\end{figure}

\subsection{Optical focusing in a cavity}
It might be of even greater relevance not to stop a pulsed molecular
beam but rather to transversely guide and collimate a subset of
continuously emerging molecules. Here we are aiming at an increased
signal for interference and spectroscopy experiments. As a first
example, we consider realizing an optical lens for these molecules
by passing them through a focused Gaussian mode of an intense laser
beam. Using a high-Q build-up cavity, intra-cavity intensities of up
to $10^4$\,W are well conceivable, which, focused to a waist of
100\,$\mu$m, would provide an optical potential depth of
$0.03$\,meV. For m=5053\,amu we see that the transverse potential
would capture all molecules with a transverse speed of up to 1\,m/s.
Hence, such a cavity enhanced laser beam close to the oven, should
increase the molecular flux at the position of the detector. Each
molecule would absorb about 200 IR photons (1064 nm) during the
passage through the focusing cavity which should be close to the
damage threshold.

To illustrate a realistic scenario we simulate the collimation of a
beam of 5000\,amu particles in a cavity enhanced laser field of
$6\times 10^4$\,W. The perturbing influence of a standing light wave
force grating would be eliminated in a ring cavity (see
Fig.~\ref{setup_cavity} a). The proposed collimation cavity is very
similar in dimensions to one used to optically trap $^{85}Rb$
atoms~\cite{Kruse2003a}. In Fig.\ref{Transversefocussing} we show
simulated trajectories of molecules with an average velocity of
$\bar{v}=50\,m/s$ and a transverse and longitudinal velocity spread
of $\pm 1\,m/s$ emanating from a square shaped source of
50$\times$50$\mu m^2$. This is the fraction of molecules which are
captured and focused by a Gaussian beam of $w_0=100\,\mu$m and
Rayleigh length of z$_R\sim$\,3\,cm, positioned at d=3\,mm above
opening of the thermal source. For a properly chosen intensity,
collimation and/or focusing can be achieved in both transverse
directions at input laser powers in the 1...10\,W range.

We therefore expect a signal gain in the forward direction by about
a factor of two, when using a single cavity and up to a factor of
four, when using two orthogonal fields. This increase is already
rather significant and  will enable interferometry experiments with
these larger perfluoroalkylated molecules~\cite{Deachapunya2007a}.

\subsection{Optical cooling in a cavity}
In an experimentally more challenging next step one can employ
cavity mediated laser cooling of these macromolecular beams. As
individual particles in are only very weakly coupled to the light
field, standard cavity cooling~\cite{Domokos2001a,Maunz2004a} is too
slow to be useful without an additional trap, but the use of
collective effects will yield a measurable result. Here we suggest a
linear confocal cavity with the axis perpendicular to both the
direction of the molecular beam and the pumping light (see
Fig.~\ref{setup_cavity} b). For a 1\,cm long cavity with a
400~\,$\mu$m waist and a line width of $\kappa \approx 2\pi\times
1\,$MHz at 1 mm above the source one would expect cooling times of
seconds for individual molecules. However, given the high number
density of $N>10^9$ perfluorinated particles in the cavity mode
volume of 1.2~mm$^3$, close to the furnace exit, one can expect a
substantial collective enhancement of the cooling effect. Additional
light scattering at intra-cavity molecules in the direction of the
mode further deepens the cooling potential and enhances cooling in
the direction of the cavity. The cavity photon number and therefore
the enhancement effect scales with $N^2$~\cite{Domokos2002a}.
Efficient cooling in a single mode can be achieved as soon as the
self-organization threshold light power $P_{T} = 1\times 10^3$ W is
reached~\cite{Asboth2005a}.

As an example, we calculate the single mode energy loss and the
associated change of the transverse velocity distribution of
perfluorinated molecules propagating perpendicular to the cavity
axis with the slowest experimentally observed velocity in
Fig.~\ref{Transversecooling}. Using two modes of the same cavity (or
two cavities) the required pump threshold corresponds to about one
Watt of input power for a cavity with a finesse $F\approx 10^3$.

\section{Conclusion}
In conclusion, we have demonstrated that even very massive molecules
can form very slow beams at useful intensities. They are very
promising for future matter-wave experiments and appear to be
accessible for a number of optical slowing and cooling experiments
with molecules in a new mass regime. Furthermore the exploration of
cryogenic cooling schemes could be interesting with these molecules.
The doping of perfluoroalkylated nanospheres into cold helium
nanodroplets would possibly be a method for enabling high resolution
spectroscopy on rather complex molecules~\cite{Stienkemeier2006a}.

This work was supported by the Austrian Science Funds FWF through
START Y177, SFB F1505 and F1512, and a Lise-Meitner fellowship (A G
M). We acknowledge support by the European Commission
(HPRN-CT-2002-00309 for E R and A S) and by the Royal Thai
government scholarship (S D). We thank William Case for fruitful
discussions.

\end{document}